\documentclass[times,twocolumn,final]{elsarticle}

\usepackage{framed,multirow}
\usepackage{algorithm}
\usepackage{algpseudocode}
\usepackage{amssymb}
\usepackage{latexsym}
\usepackage{url}
\usepackage{amsfonts} 
\usepackage{amsmath}
\usepackage{amsthm}
\usepackage{pifont}
\usepackage{natbib}
\usepackage{geometry}
\usepackage{graphicx}
\usepackage{color}

\usepackage{algorithmicx}
\usepackage{setspace}
\usepackage{graphicx,psfrag,epsf}
\usepackage{booktabs}
\usepackage{algorithm}
\usepackage{algpseudocode}

\floatname{algorithm}{Algorithm}
\usepackage{url, amssymb, amsfonts}
\usepackage[colorlinks = true,
linkcolor = blue,
urlcolor  = blue,
citecolor = blue,
anchorcolor = blue]{hyperref}
\usepackage{graphicx}
\usepackage{amsmath}
\usepackage{paralist}
\usepackage{psfrag}
\usepackage{natbib}
\usepackage{url} 
\usepackage{setspace}
\usepackage{lscape}
\usepackage{color,amssymb}
\usepackage{amsthm}
\usepackage{dcolumn}
\usepackage{indentfirst, verbatim, float}
\theoremstyle{plain}
\usepackage{enumerate}

\newtheorem{definition}{Definition}
\newtheorem{theorem}{Theorem}
\newtheorem{lemma}[theorem]{Lemma}

\usepackage{color}

\newcommand{\EE}{\mathbb{E}}   
\newcommand{\PP}{Prob}   
\begin{document}

\begin{frontmatter}
\title{Discovering the Signal Subgraph: An Iterative Screening Approach on Graphs}
\author[1]{Cencheng Shen\corref{cor1}}
\ead{shenc@udel.edu}
\author[2]{Shangsi Wang}
\ead{swang127@jhu.edu}
\author[3]{Alexandra Badea}
\ead{alexandra.badea@duke.edu}
\author[2]{Carey E. Priebe}
\ead{cep@jhu.edu}
\author[4]{Joshua T. Vogelstein}
\ead{jovo@jhu.edu}

\address[1]{Department of Applied Economics and Statistics, University of Delaware}
\address[2]{Department of Applied Mathematics and Statistics, Johns Hopkins University}
\address[3]{Center for In Vivo Microscopy, Duke University}
\address[4]{Department of Biomedical Engineering and Institute for Computational Medicine, Johns Hopkins University}
\cortext[cor1]{Corresponding author. Cencheng Shen and Shangsi Wang contribute equally to this work.}

\begin{abstract}
Supervised learning on graphs is a challenging task due to the high dimensionality and inherent structural dependencies in the data, where each edge depends on a pair of vertices. Existing conventional methods are designed for standard Euclidean data and do not account for the structural information inherent in graphs. In this paper, we propose an iterative vertex screening method to achieve dimension reduction across multiple graph datasets with matched vertex sets and associated graph attributes. Our method aims to identify a signal subgraph to provide a more concise representation of the full graphs, potentially benefiting subsequent vertex classification tasks. The method screens the rows and columns of the adjacency matrix concurrently and stops when the resulting distance correlation is maximized. We establish the theoretical foundation of our method by proving that it estimates the true signal subgraph with high probability. Additionally, we establish the convergence rate of classification error under the Erdos-Renyi random graph model and prove that the subsequent classification can be asymptotically optimal, outperforming the entire graph under high-dimensional conditions. Our method is evaluated on various simulated datasets and real-world human and murine graphs derived from functional and structural magnetic resonance images. The results demonstrate its excellent performance in estimating the ground-truth signal subgraph and achieving superior classification accuracy.
\end{abstract}
\begin{keyword}
iterative screening, distance correlation, graph classification
\end{keyword}
\end{frontmatter}

\section{Introduction}\label{sec:introduction}
The analysis of graph structure is critical in various big data fields, including neuroscience, internet mapping, and social networks \citep{otte2002social,newman2002, bullmore2011brain, vogelstein2013graph,ShenVogelsteinPriebe2016, OGBData}. Due to the large size of graphs in practice, such as in social networks and raw neuroimages, it is often necessary to use smaller subgraphs from the observed graphs. Moreover, the selected subgraph should maintain or improve subsequent inference. For example, identifying a subset of brain regions from brain imaging to better predict the phenotype of interest in each subject.

The statistical problem of feature reduction and dimension selection has been extensively studied, with well-known methods such as Lasso \citep{Tibshirani1996}, adaptive Lasso \citep{ZouHastie2006}, Dantzig selector \citep{CandesTao2007}, sure independence screening \citep{fan2008sure,li2012feature}, among others. These methods have specific objectives, such as sparsity and the recovery of ground-truth. Among them, the screening method is known for its computational efficiency and model-free nature and are commonly used for high-dimensional data \cite{Zhu2011model}, making it a suitable candidate for graph data. 

However, dimension reduction for graph data presents unique challenges because of its high-dimensionality and the unique structure of the $n \times n$ adjacency matrix. To that end, this paper proposes an iterative screening method on graph data, which utilizes distance-based correlation and independence screening in an iterative manner to estimate the signal subgraph. During each iteration, we define the feature for each vertex using the adjacency of the reduced graphs, compute a distance-based correlation between the feature and the label of interest $Y$, and discard vertices with small correlations. This process is repeated recursively on the reduced graphs from previous iterations, yielding a smaller set of vertices each time until an estimated signal subgraph is selected for output. 

The proposed method is straightforward to use and implement. We provide theoretical results that demonstrate the method's ability to identify the true signal vertices with high probability. Additionally, our approach guarantees asymptotically optimal classification performance under the Erdos-Renyi random graph model, outperforming the use of the entire graph in specific high-dimensional settings. Simulation results showcase the superior performance of the proposed method, including improved prediction accuracy when compared to conventional non-iterative screening approaches or using the full graph, as well as accurate estimation of the ground-truth signal subgraph. Furthermore, the paper demonstrates the method's applicability to MRI brain graphs for studying site effects and sex differences in brain imaging analysis. It successfully identifies regions that minimize validation error, thereby pinpointing potential regions of interest for practitioners. 

\section{Preliminaries}
\label{sec:main}
\subsection{Setting and Notations}
Given $m$ observed graphs $\{A_i, i=1,\ldots,m\}$ with a shared vertex set $V=[n]$, we shall slightly abuse the notation and also denote $A_i \in \mathbb{R}^{n \times n}$ as the adjacency matrix of the graph. The graphs can be weighted or unweighted, and directed or undirected. Given any subset of vertices $U\subseteq V=[n]$, the reduced adjacency matrix is denoted by ${A_{i}(U)}$, which is the subgraph using $U$. Furthermore, each graph is associated with a label of interest $\{Y_i \in \mathbb{R}, i=1,\ldots,m\}$.

In the classical statistical pattern recognition setting, the pairs of observations $\{(A_i,Y_i)\}_{i=1}^m$ are independent and identically distributed pairs according to a distribution $F_{A,Y}$ \citep{devroye2013probabilistic}, that is 
\[(A_1,Y_1),(A_2,Y_2),(A_3,Y_3),...,(A_m,Y_m) \overset{i.i.d.}{\sim} F_{A,Y}\]
for some true but unknown joint distribution, where $A$ denotes the underlying random variable of $\{A_i\}$ and $Y$ represents the underlying random variable of $\{Y_i\}$. Moreover, we denote $g(\cdot)$ as a given classifier, and the resulting classification error as \[L(g) = \PP(g(\cdot) \neq Y).\] In addition, we denote the Bayes optimal classifier as $g^*(\cdot)$, so a classifier is asymptotically optimal if and only if $L(g) \rightarrow L(g^*)$.

It is often the case that $Y$ depends only on a small portion of $A$, which motivates the need for a definition of signal subgraph and signal vertices.
\begin{definition} For any subset of vertices $U \subset V=[n]$, denote the induced subgraph of $U$ by $A(U)$, and denote the subgraph removing all edges in $A(U)$ as $A(U^{-})$. The set of \textbf{\textit{signal vertices}} $S$ is defined to be the minimal subset of vertices $U$, such that $A(U^{-})$ is independent of $Y$, that is
\[S = \arg\min_{U} |U| \text{ , subject to } A(U^{-}) \perp Y,\]
where the notation $\perp$ means independence between the subgraph and the label. The induced graph on the signal vertices $S$ is called the \textbf{\textit{signal subgraph}}.
\end{definition}
If the graph $A$ is independent of $Y$, there is no signal in the graph, resulting in $S=\varnothing$. If all vertices in $A$ are incident on at least one edge which is dependent on $Y$, then $S=V$. The signal subgraph from this definition may not be unique, but one such subgraph suffices, because the subsequent classification is always asymptotically optimal as shown in Section~\ref{sec:theory}. 

\subsection{Distance Correlation}

The distance correlation is a measure that can detect all types of dependencies between two random variables, given sufficient sample size \citep{szekely2007measuring}. To compute the sample distance correlation, two pairwise distance matrices are transformed and multiplied using a Hadamard product. The sample distance correlation is asymptotically $0$ if and only if the two underlying random variables are independent. For more detailed mathematical information about the distance correlation and its population definition, see the Appendix.

The distance correlation is a computationally efficient method \citep{DCorFast}, and has been shown to be equivalent to kernel correlation \cite{DCorKernel}. It has been used for various inference tasks \citep{Wang2015,Pitsillou2018, DCorHD,DCorTemporal,DCorMANOVA} not limited to screening. This paper also utilizes a local version of the distance correlation called the multiscale graph correlation (MGC), which improves testing power against nonlinear dependencies \citep{MGC,MGCDCor,MGCGraph,DCorGraph}.

\section{Main Method}
\label{sec:mainalg}
The proposed iterative vertex screening algorithm consists of three steps: extracting features within each reduced graph, computing distance-based correlation between the feature and the label of interest, then iteratively reducing the graph size by a factor $\delta \in (0,1)$ through discarding vertices with low correlation. The algorithm outputs a set of vertices $\hat{S}$ that estimates the true signal vertices $S$. Algorithm~\ref{alg:ivs} presents the proposed iterative method, while Algorithm~\ref{alg:vs} describes a conventional screening method used as a benchmark in the simulations.

The first step computes a feature vector for each vertex within the reduced graph. At each iteration $k$, denote the current reduced vertex set as $U_{k}$, we use $A_i(U_{k})[u,\cdot]$ as the $i$th feature of vertex $u$, i.e., the $u$th row of adjacency matrix $A_i$ restricted to the vertex set $U_{k}$. The second step computes a dependency measure $\beta(u)$ between $\{A_i(U_{k})[u,\cdot]\}_{i=1}^m$ and $\{Y_i\}_{i=1}^m$ for each vertex $u$. Either distance correlation (Dcor) or multiscale graph correlation (MGC) can be used for $\beta(u)$ (or one could use any other correlation, like the traditional Pearson correlation, kernel correlation), denoted by
\begin{align*} 
\beta(u) = & Dcor(\{(A_i(U_{k})[u,\cdot] , Y_i)\}_{i=1}^m ), \text{ or }  \\ 
\beta(u) =& MGC(\{(A_i(U_{k})[u,\cdot] , Y_i)\}_{i=1}^m). 
\end{align*}
Then the vertices are sorted based on the magnitude of their $\beta(u)$ values, and a critical value $t$ is determined via percentile. Vertices with $\beta(u)$ values below $t$ are discarded, and the remaining vertices form the vertex set $U_{k+1}$ for the next iteration, i.e., 
\begin{align*} 
U_{k+1} = \{u \in U_{k} | \beta(u) > t\}.
\end{align*}
The choice of $\delta$ is at the discretion of the user and depends on their desired level of conservatism regarding vertex removal. For instance, selecting $\delta=0.5$ results in the removal of half of the vertices at each iteration, striking a balance between running time and performance, particularly for large datasets. Conversely, when dealing with moderate sample sizes, opting for a smaller $\delta$ value, such as $\delta=0.05$, leads to the removal of only a few vertices in each step. This choice represents a more cautious approach, requiring additional computational time but potentially yielding greater accuracy. The simulations were conducted to compare the performance of both choices. 

The iteration continues until only one vertex remains, or until the desired size of the output vertex set is reached. The algorithm calculates the distance correlation between the graph feature ${A_i(U_k)}$ and the label vector for each subgraph produced from each iteration. The final output is the set of vertices that maximizes the correlation. An alternative approach is to use cross-validation to select the subgraph with the best leave-one-out prediction error, which is computationally more expensive but has similar empirical performance, as demonstrated in Section~\ref{sec:exp} and Section~\ref{sec:app}.

\begin{algorithm}
	\caption{Iterative Vertex Screening}
	\label{alg:ivs}
	\begin{algorithmic}[1]
		\Require{$\{(A_i,Y_i)\}_{i=1}^m$ , $\delta \in (0,1)$}{}
		\State Set $k=1$, and $U_k = V$
		\While{$|U_k|>1$}
		\For{$u \in U_k$ }
		\State $X_i=A_{i}(U_{k})[u,\cdot]$
		\State $\beta(u) = Dcor(\{X_i,Y_i\}_{i=1}^m )$
		\EndFor
		\State Set $t$ be the $\delta$ quantile among $\{\beta(u), u \in U_k\}$
		\State Set $U_{k+1} = \{u \in U_k|\beta(u) > t\}$
        \State Set $k = k+1$
		\EndWhile
        \State $k^{*}=\arg\max_{k} Dcor(\{(A_{i}(U_{k}),Y_i)\}_{i=1}^{m})$
		\State Output the signal vertices $\hat{S} = U_{k^{*}}$.
	\end{algorithmic}
\end{algorithm}

\begin{algorithm}
	\caption{Conventional Screening Applied to Graphs}
	\label{alg:vs}
	\begin{algorithmic}[1]
		\Require{$\{(A_i,Y_i)\}_{i=1}^m$ and $c \in [0,1]$}{}
		\For{$u \in V$ }
		\State $X_i=A_{i}[u,\cdot]$
		\State $\beta(u) = Dcor(\{X_i, Y_i\}_{i=1}^m )$
		\EndFor
		\State $\hat{S} = \{u \in V| \beta(u) > c\}$. 
	\end{algorithmic}
\end{algorithm}

\section{Theoretical Properties}
\label{sec:theory}
To establish the theoretical properties, we make the following assumptions:
\begin{itemize}
\item The number of vertices in the ground-truth signal vertices $|S|=p$ is fixed.
\item $\hat{S}$ is estimated using Algorithm~\ref{alg:ivs} and distance correlation, with the output vertex set satisfies $|\hat{S}| \geq p$.
\item The graph adjacency matrix $A$ follows the Erdos-Renyi random graph model \citep{erdds1959random} and has a bounded Frobenius norm.
\item The Bayes plug-in classifier $g(\cdot)$ is used.
\end{itemize}

Under the above assumptions, the estimated signal vertices include the truth with high probability:
\begin{theorem} 
\label{thm1}
There exist two positive constants $c_1, c_2$ and some $0<\gamma < 1/2$ such that
\begin{align*}
\PP(S & \subset \hat{S} )  > \\
&1 - O(p \exp(-c_1 m ^{1-2\gamma}) + mp \exp(-c_2 m^{\gamma})).
\end{align*}
In particular, $\PP(S \subset \hat{S} ) \rightarrow 1$ as $m \rightarrow \infty$.
\end{theorem} 

Next, we establish the convergence rate of the classification error using the estimated subgraph, where $g(\hat{S})$ represents the plug-in classifier utilizing the estimated signal subgraph.
\begin{theorem}
\label{thm2}
	With high probability, $L(g(\hat{S})) - L(g^*)$ is bounded by $\epsilon$. Specifically, there exist four positive constants $c_1$, $c_2$, $c_3$, $c_4$, such that
	\begin{align*}
		&\PP(L(g(\hat{S})) - L(g^*) < \epsilon)  \geq  \\
        & \qquad 1 -2(E(\hat{S})+1)\exp \left( \frac{-m c_4 \epsilon^2}{(2E(\hat{S})+\sqrt{2c_4})^2} \right)\\
        & \qquad - c_3 (p \exp(-c_1 m ^{\frac{1}{3}}) + mp \exp(-c_2 m ^{\frac{1}{3}})),
	\end{align*}
	where $E(\hat{S})$ denotes the expected number of edges in the estimated signal subgraph. 
\end{theorem}

Therefore, the classification performance using the estimated signal subgraph is asymptotically optimal. In contrast, using the whole graph for classification is suboptimal when the size of the graph is as large as the fourth root of the sample size.
\begin{theorem}
\label{thm3}
As sample size $m$ approaches infinity, it holds that 
\begin{align*}
L(g(\hat{S})) \rightarrow L(g^*).
\end{align*}
Moreover, when $n=O(m^{\frac{1}{4}})$, for sufficiently large $n$ and $m$, it holds that
\begin{align*}
L(g(\hat{S}))&<L(g(V)) \\
\mbox{ and } L(g(V)) &> L(g^*).
\end{align*}
\end{theorem}
The results suggest that using the estimated signal subgraph via iterative screening can be expected to perform better than using the whole graph, when the size of the signal subgraph is fixed and the number of observed graphs is comparable to the size of the whole graph. This setting is illustrated in the top panel of Figure~\ref{fig:vscl} with $n=200$, $m=300$, and $|S|=20$, and is also observed in the experiments in Section~\ref{sec:exp} where the estimated subgraph yields better classification performance. All proofs and additional mathematical background are provided in the appendix.

\section{Simulations}
\label{sec:exp}
\subsection{Signal Subgraph Estimation}
We generate 100 Erdos-Renyi graphs (ER) from two classes. The graph is generated by $A|Y=y \sim ER(P_y)$ with $y \in \{0,1\}$ and
\[P_y = \begin{bmatrix} P_y\times\mathbf{1}_{20\times20} &  0.2 \times \mathbf{1}_{20\times180} \\ 0.2 \times \mathbf{1}_{180\times20} & 0.3 \times \mathbf{1}_{180\times180} \end{bmatrix}, \]
where $P_0 = 0.3$ and $P_1 = 0.4$. Namely, the graph contains 200 vertices, out of which only the first 20 vertices are signal vertices containing information to separate $y=0$ from $y=1$. More information on the Erdos-Renyi model is provided in the appendix.

\begin{figure}[!ht]
	\centering
	\includegraphics[width = 3.0in]{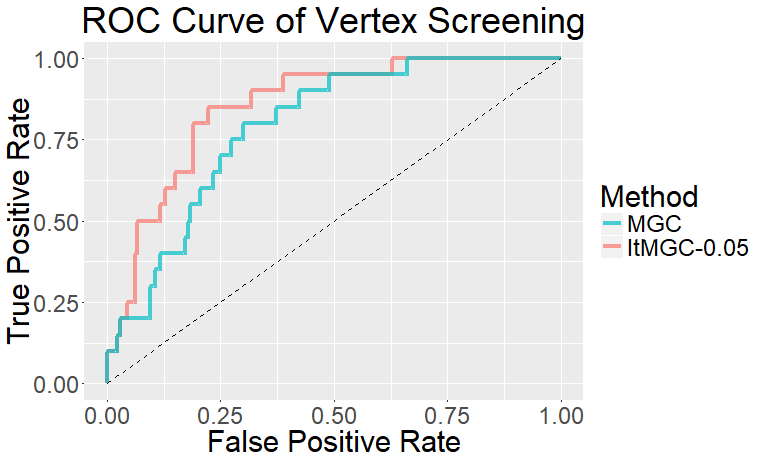}
	\caption{The figure shows the Receiver operating characteristic (ROC) of the iterative vertex screening and conventional screening using MGC. It is evident that the iterative vertex screening performs significantly better.}
	\label{fig:vs}
\end{figure}

We estimate the subgraph using various screening methods, including the conventional screening with Dcor and MGC, iterative screening with Dcor and MGC at $\delta=0.5$ and $\delta=0.05$ respectively, and screening with canonical correlation analysis (CCA) \citep{hotelling1936relations} and RV coefficient (RV) \citep{robert1976unifying}. Since the actual size of the signal subgraph was known to be 20, we output the estimated subgraph at the same size and calculated the true positive rate. The ROC curve is shown in Figure \ref{fig:vs}, while Table \ref{tb:vs} reports the AUC and runtime for each approach. We observe that Dcor and MGC outperform CCA and RV, and terative screening improves the performance over conventional screening. Furthermore, iterative screening with $\delta=0.05$ yields better results than iterative screening with $\delta=0.5$ at the cost of a longer running time.

\begin{table}[!ht]
	\centering
 \caption{This table presents the mean and standard error of the area under the curve (AUC) and running time of the eight methods based on $100$ replicates. Iterative vertex screening outperforms conventional screening. The running times were measured using MATLAB R2022a on a standard laptop equipped with a 16-core Intel CPU and 16GB of memory.}
    	\begin{tabular}{|l|l|l|}
		\hline
		Method & AUC & Time (sec)  \\\hline 
        ItDcor-0.05 & 0.8705 (0.0113) & 18.50 (1.35)\\\hline
        ItDcor-0.50 & 0.8655 (0.0094) & 2.03 (0.17)\\\hline
        ItMGC-0.05 & 0.8720 (0.0122) & 967.42 (17.73)\\\hline
        ItMGC-0.50 & 0.8625 (0.0106) & 120.16 (7.32)\\\hline
        Dcor & 0.8554 (0.0056) & 1.23 (0.22) \\\hline
        MGC & 0.8555 (0.0057) & 38.44 (1.720) \\\hline
        RV & 0.8506 (0.0077) & 2.12 (0.10) \\\hline
        CCA & 0.5353 (0.0080) & 0.92 (0.04) \\\hline
	\end{tabular}
	\label{tb:vs}
\end{table}

\subsection{Classification Accuracy}
\begin{figure}[!ht]
	\centering
	\includegraphics[width = 3.4in ]{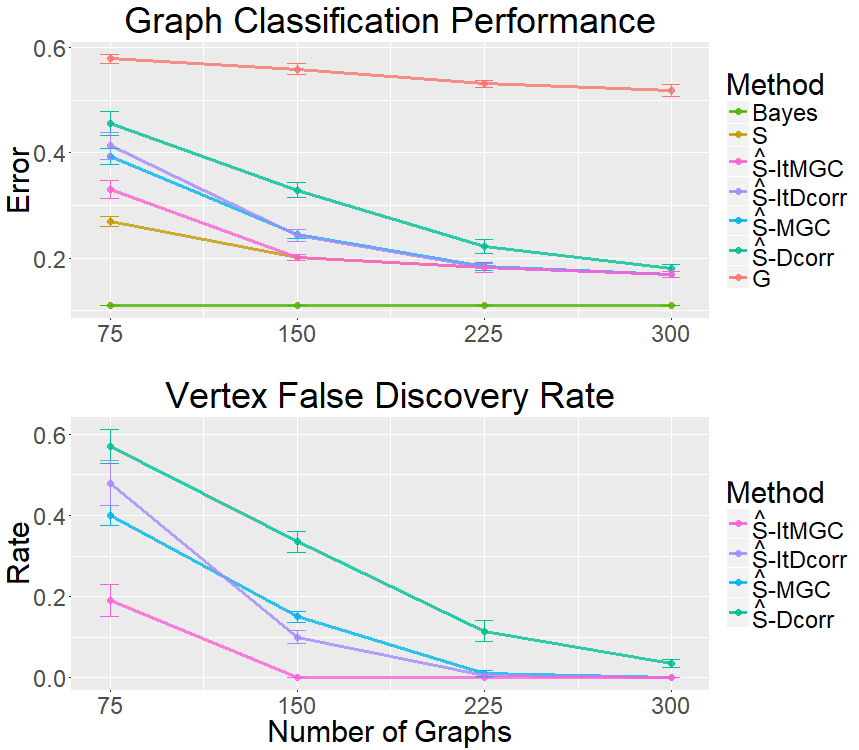}
	\caption{In the top panel, we compare seven classifiers: the Bayes optimal classifier, Bayes plug-in using $S$, Bayes plug-in using $\hat{S}$ estimated by iterative Dcor or MGC, Bayes plug-in using $\hat{S}$ estimated by Dcor or MGC, and Bayes plug-in using $G$. The bottom panel displays the false discovery rate in estimating the signal vertices. The mean results are reported using $100$ independent simulations, and the error bars represent two times the standard deviation. Note that the bottom panel only considers four methods that estimate the signal vertices, omitting the Bayes classifier, the Bayes plug-in using $S$, and the Bayes plug-in using $G$. These three methods do not involve the estimation of the signal subgraph and are not applicable to the bottom panel.}
	\label{fig:vscl}
\end{figure}

Here we investigate the classification performance using the estimated signal subgraph. We consider a $3$-class classification problem using the Erdos-Renyi model, and generate $A|Y=y \sim ER(P_y)$ with $y \in \{0,1,2\}$ and
\[P_y = \begin{bmatrix} P_y \times \mathbf{1}_{20\times20} &  0.2 \times\mathbf{1}_{20\times180} \\ 0.2 \times \mathbf{1}_{180\times20} & 0.3 \times \mathbf{1}_{180\times180} \end{bmatrix}, \]
where 
\[ P_y =
\begin{cases}
0.4       & \quad \text{if } y=0,\\
0.3  & \quad \text{if } y=1,\\
0.5  & \quad \text{if } y=2.\\
\end{cases}
\]
Each graph has 200 vertices, with the first 20 vertices designated as signal vertices. We consider the Bayes plug-in error $L(g(\hat{S}))$ using conventional Dcor and MGC screening as well as iterative vertex screening using Dcor and MGC, respectively. We then compare the results to $L(g)$, $L(g^*)$, and $L(g(S))$, representing the plug-in error using all vertices, the Bayes optimal error, and the plug-in error using the true signal vertices. Figure \ref{fig:vscl} illustrates the classification error and false discovery rate in detecting the signal vertices.

The results indicate that using the estimated signal subgraph leads to better classification performance compared to using the entire graph. MGC performs better than Dcor, and the iterative approach outperforms the conventional method. Moreover, the screening method accurately recovers the actual signal subgraph after $m>300$, and the classification error approaches the Bayes optimal. Since this experiment has a comparable design to the previous one, CCA or RV are not considered as they have inferior performance.

\begin{figure}[!ht]
	\centering
	\includegraphics[width = 3.4in]{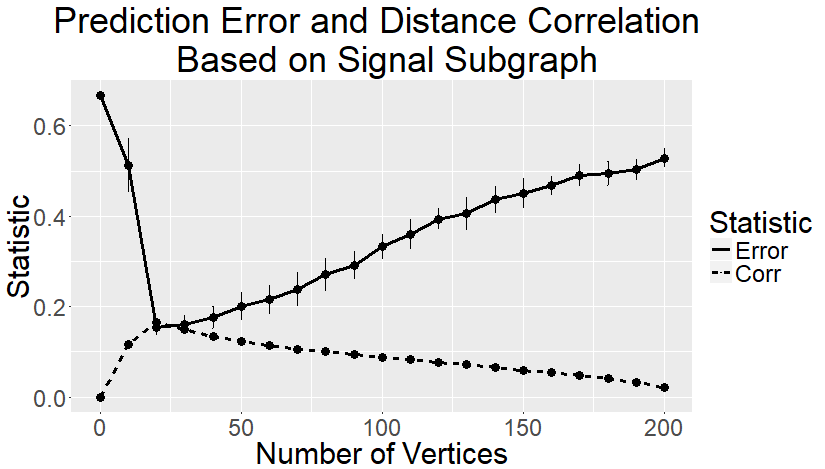}
	\caption{The figure shows the prediction error and distance correlation for subgraphs of varying sizes produced by the iterative Dcor screening algorithm. The algorithm produces a subgraph with 20 vertices, which matches the actual size of the signal subgraph and also results in the lowest prediction error. The mean results are reported using $100$ independent simulations, and the error bars represent two times the standard deviation. }
    \label{fig:vssize}
\end{figure}

Since the size of $S$ is typically unknown in practice, our next simulation evaluates the stopping criterion in Algorithm~\ref{alg:ivs}, which outputs the estimated subgraph that maximizes the distance correlation. Figure \ref{fig:vssize} illustrates that the criterion performs as expected in this experiment at $m=300$: $\hat{S}$ with $20$ vertices indeed maximizes the distance correlation, corresponds to the actual number of true signal vertices, thus effectively minimizes the prediction error.

Therefore, this figure showed two points: first, it is important to estimate the signal subgraph, as a smaller graph can leads to significantly better classification performance; second, our iterative screening algorithm worked as intended, which stopped at maximum correlation and successfully estimates the best signal subgraph in this case. 

\section{Study on Brain Imaging}
\label{sec:app}
\subsection{Site and Sex Prediction With Human Brain} 
Our objective is to predict the sex and site of each individual based on functional magnetic resonance image (fMRI) graphs \citep{ogawa1990brain}. We utilized two datasets, SWU4 \citep{qiuswu4} and HNU1 \citep{wenghnu1}, which include $467$ and $300$ subjects, respectively. Each individual's fMRI scan is registered to the MNI152 template using the Desikan atlas, which has $70$ regions \citep{desikan2006automated}. The graphs are created using the NeuroData's MRI Graphs pipeline\footnote{\url{https://github.com/neurodata/ndmg}}, a popular tool for processing and representing brain images.

We perform a leave-one-subject-out signal subgraph estimation and prediction process. We use the site information as the label vector and apply iterative vertex screening via distance correlation to all graphs, except for one that is left out. Next, we utilize $9$-nearest-neighbor to predict the site of the left-out subject. We repeat this process for each subject, calculate the leave-one-out classification error, and repeat it for the sex information as the label vector. Note that the performance is robust against different nearest-neighbor parameters, and in this case, we selected the nearest odd integer to $\log_2$ of the sample size, which resulted in choosing a $9$-nearest-neighbor.

\begin{figure}[!ht]
	\centering
	\includegraphics[width=3.5in]{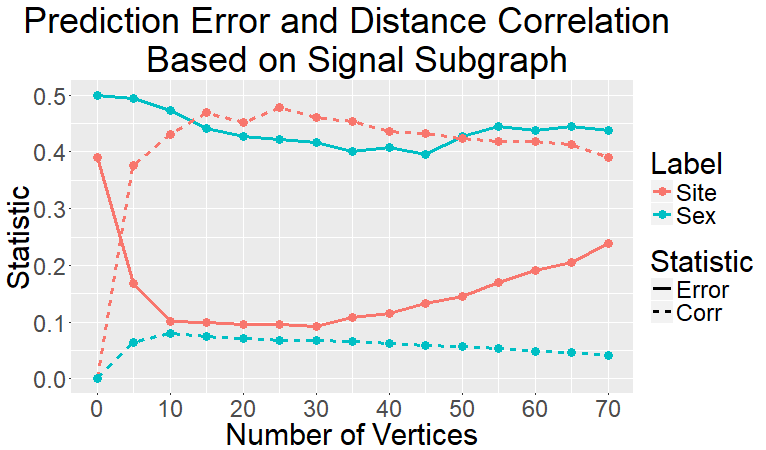}
	\caption{The figure shows the leave-one-subject-out prediction error and distance correlation at various sizes of estimated subgraph. The data set combines two studies, SWU4 and HNU1, and we perform a leave-one-subject-out screening and classification to identify brain regions that are significant for differentiating sex or site. }
	\label{fig:study}
\end{figure}

Figure \ref{fig:study} illustrates the prediction error and distance correlation in relation to the varying size of the estimated subgraph produced by Algorithm~\ref{alg:ivs}. The red lines represent site classification, while the blue lines denote sex classification. In terms of sex differences, we observe that there is no prominent signal in the data, as neither the distance correlation nor the classification error are notably superior. For site classification, the iterative screening algorithm produces a subgraph containing $30$ vertices, which maximizes the distance correlation and also minimizes the classification error.

\begin{figure}[!ht]
	\centering
	\includegraphics[trim={0cm 4.0cm 0 4.5cm},clip,width = 3.0in]{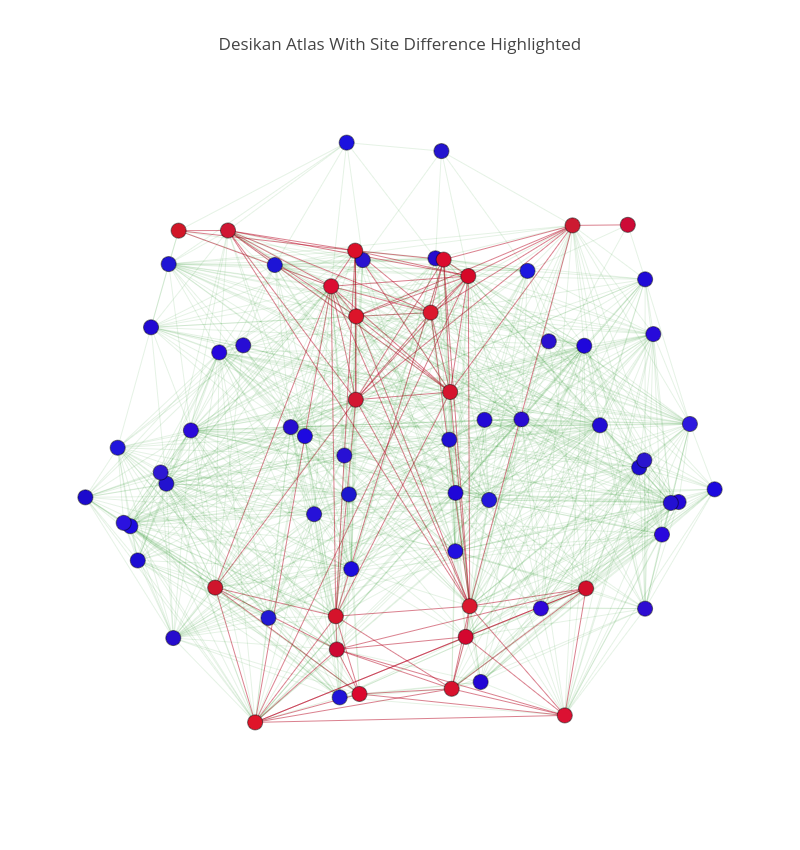}
	\caption{The figure displays the Desikan atlas, with highlighted brain regions that are substantially associated with site. The $11$ matched brain regions identified in Table \ref{tb:match} are shown in red and are spatially adjacent to each other.}
	\label{fig:brain}
\end{figure}

The estimated signal vertices provide additional insight into the graph structure. Specifically, the vertices chosen for site difference are exactly matched across the left and right hemispheres. If we consider the $35$ paired regions in the Desikan atlas, we can categorize the pairs based on whether both regions are among the $30$ estimated signal vertices or not. The outcome is presented in Table \ref{tb:match}. The regions with large distance-based correlations are significantly matched. based on a chi-square test yielding a p-value of $0.002$. The $11$ left-right hemisphere matched regions include caudal anterior cingulate, corpus callosum, cuneus, fusiform, lateral occipital, lingual, parsorbitalis, precuneus, rostral anterior cingulate, rostral middle frontal gyrus, and superior frontal gyrus, as shown in Figure \ref{fig:brain}.
\begin{table}[!ht]
	\centering
 \caption{The number of left-right hemisphere matched regions with large or small distance-based correlations.}
	\begin{tabular}{|l|l|l|}
		\hline
		Number of Pairs& Right-Large & Right-Small \\\hline
		Left-Large & 11 & 1 \\\hline
		Left-Small &  7 & 16 \\\hline
	\end{tabular}
	\label{tb:match}
\end{table}

\subsection{Sex Difference in Mouse Brain} 
Structural magnetic resonance imaging has provided insight into the genetic basis of mouse brain variability by examining the relationship between volume covariance and genotypes \citep{badea2009genetic}. With high-resolution diffusion tensor imaging and tractography, we can now investigate the underlying bases for structural connectivity patterns \citep{calabrese2015diffusion}, in relationship with genotype and sex. Based on MRI and conventional Nissl histology, we scanned and registered $55$ mouse brains (pooled genotypes) into the space of a minimum deformation template, aligned to Waxholm space \citep{johnson2010waxholm}. The atlas labels were propagated onto the template and, subsequently, onto each individual brain using ANTs \citep{avants2011reproducible}. We employed DSI Studio \citep{yeh2013deterministic} to estimate tract-based structural connectivity for each brain, which was then represented as a graph with $332$ vertices, $166$ per hemisphere. Of the $55$ mice, $32$ are male, and $23$ are female.

\begin{figure}[!ht]
	\centering
	\includegraphics[width = 3.0in]{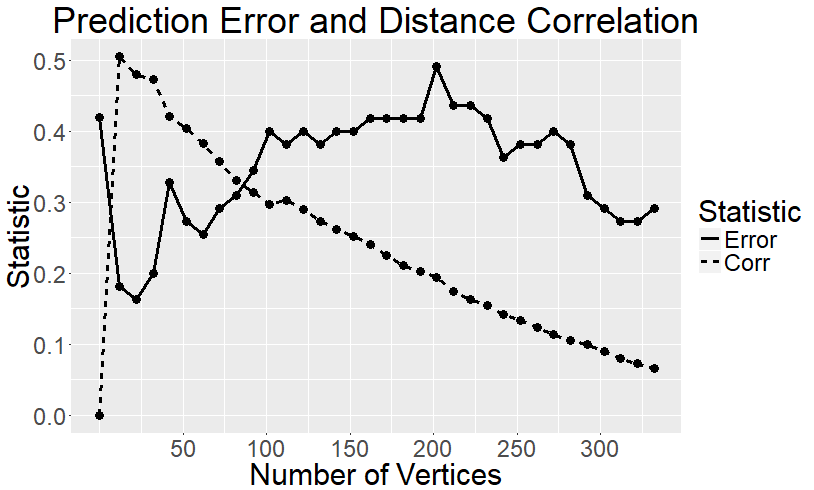}
	\caption{The figure depicts the prediction error and distance correlation from various sizes of the estimated signal subgraph for mouse sex classification. This demonstrates that a smaller signal subgraph (with a size of 10) yields a better classification error compared to the full graph. Additionally, our designed iterative screening algorithm successfully identifies the signal subgraph, which maximizes the correlation. }
	\label{fig:mouse}
\end{figure}

Similarly, we conduct a leave-one-out evaluation using an iterative vertex screening to estimate the signal subgraph, followed by a $9$-nearest-neighbor classifier to predict the left-out sample based on the estimated signal subgraph. Figure \ref{fig:mouse} demonstrates the prediction error and distance correlation when using the iterative screening algorithm. Despite the small sample size and fluctuating prediction error, the screening method outputs a signal subgraph of size $10$, which results in a near-optimal classification error of $0.18$. The estimated signal vertices include a thalamic component and the periaqueductal gray, which play an important role in driving sexually dimorphic mouse brain development \citep{spring2007sexual, lerch2013sex}.

\section{Conclusion}
In summary, we developed an iterative vertex screening methodology to estimate the signal subgraph of interest and successfully applied the method in simulations and real data. Utilizing distance correlation and multiscale graph correlation lends strong interpretability to our methods. Given the existence of signal vertices where each vertex is dependent on the graph-level attributes (while non-signal vertices are independent), Theorem~\ref{thm1} suggests that the signal subgraph can be recovered with probability converging to $1$ as the number of graphs increases. Furthermore, Theorems~\ref{thm2} and~\ref{thm3} suggest that subsequent classification using the signal vertices can be asymptotically Bayes optimal, and in certain cases (depending on the relationship between $n$ and $m$), better than utilizing the full graph.

We shall emphasize that the proposed method is essentially a dimension reduction technique, and one could use any subsequent classifier, not just Bayes plug-in. Therefore, Theorems~\ref{thm2} and~\ref{thm3} should be viewed as providing theoretical guarantees under a simple classifier case. Nevertheless, Theorem~\ref{thm1} is a very general result that focuses solely on dimension reduction. Intuitively, excluding independent vertices is expected to benefit many subsequent tasks beyond classification. Previous research has demonstrated that dependence measures can lead to better interpretability and improvements in complex machine learning architectures \citep{Guo2024, Zhen2022}.

The experiments and theories provide strong evidence that the iterative approach effectively and accurately estimates the signal subgraph, resulting in better performance for subsequent classification compared to conventional screening methods. It is important to emphasize once again that our method requires multiple graph datasets with a common set of vertices, and it has been shown that incorporating more graphs can enhance subsequent vertex classification \citep{GEEFusion}. If there are multiple graphs available but the vertices are not matched, then our method is not applicable. However, if part of the vertex set is matched across graphs, our method can still be applied to the matched vertex subset. In cases where the vertices are matched but the actual correspondence is unknown, graph matching techniques may be applied first \citep{LyzinskiFishkindPriebe2014, LyzinskiFishkindPriebe2016}.

\section*{Acknowledgement}
\addcontentsline{toc}{section}{Acknowledgment}
The authors gratefully acknowledge support from the Defense Advanced Research Projects Agency's (DARPA) GRAPHS program through contract N66001-14-1-4028, the DARPA SIMPLEX program through contract N66001-15-C-4041, the DARPA D3M program through contract FA8750-17-2-0112, the DARPA Lifelong Learning Machines program through contract FA8650-18-2-7834, the National Science Foundation awards DMS-1921310 and DMS-2113099, and the National Institutes of Health through R01 MH120482, K01 AG041211, R56 AG057895, P41 EB015897 and S10 OD010683. The authors would like to thank Dr. Daniel S.~Margulies for useful feedback, and Dr. Carol Colton for her advice on the mouse experiments.

\bibliographystyle{model2-names}
\bibliography{reference}

\clearpage
\onecolumn
\setcounter{figure}{0}
\setcounter{section}{0}
\renewcommand{\thealgorithm}{C\arabic{algorithm}}
\renewcommand{\thefigure}{E\arabic{figure}}
\renewcommand{\thesection}{A.\arabic{section}}
\renewcommand{\thesubsection}{\thesection.\arabic{subsection}}
\renewcommand{\thesubsubsection}{\thesubsection.\arabic{subsubsection}}
\pagenumbering{arabic}
\renewcommand{\thepage}{\arabic{page}}

\bigskip
\begin{center}
{\large\bf APPENDIX}
\end{center}

\section{Technical Preliminaries}
\subsection{Distance Correlation}

Given sufficient sample size, the distance correlation \citep{szekely2007measuring} is able to detect all types of dependencies between two random variables. The population distance covariance $Dcov(X,Y)$ can be defined via either the characteristic functions or Euclidean distance as:
\begin{align*}
Dcov(X,Y) &= \frac{1}{c_p c_q} \int_{} \int_{ } \frac{|\phi_{X,Y}(s,t) - \phi_{X}(s) \phi_{Y}(t)|^2 }{\|s\|^{1+p} \|t\|^{1+q}} dt ds \\
&=  \EE(\|X-X'\|\|Y - Y'\|) + \EE(\|X-X'\|)\EE(\|Y -Y'\|) \\
& - 2 \EE(\|X-X'\|\|Y - Y''\|),
\label{eq:dcov}
\end{align*}
where $\phi_{X,Y}$, $\phi_{X}$ and $\phi_{Y}$ are characteristic functions of $(X,Y)$, $X$ and $Y$ respectively, $c_p$ and $c_q$ are constants, and $(X,Y), (X',Y'), (X'',Y'')$ are independent and identically distributed as $F_{XY}$. The population distance correlation $Dcor(X,Y)$ between $X$ and $Y$ is
\[Dcor(X,Y) = \frac{Dcov(X,Y)}{ \sqrt{Dcov(X,X) Dcov(Y,Y)}} \in [-1,1],\]
which equals $0$ if and only if $X$ and $Y$ are independent. Then the sample distance correlation is defined via taking a Hadamard product between sample distance matrices. The sample version converges to the population, thus asymptotically $0$ if and only if independence. 

\subsection{Graph Classification}
We introduce the binary classification setting of predicting the label $Y \in \{0,1\}$ using graph $A$. This set-up serves as the basis for Section~\ref{sec:theory} and the simulations. The network model under consideration is the inhomogeneous Erdos-Renyi (ER) random graph model \citep{erdds1959random}, which allows edges to have different probabilities and generates a family of distributions on undirected graphs. The ER model can also be viewed as a stochastic block model with each block containing only one vertex.
\begin{definition} Inhomogeneous Erdos-Renyi model (ER). A random adjacency matrix $A$ is said to follow an inhomogeneous Erdos-Renyi random graph model with edge probability matrix $P \in [0,1]^{n\times n}$, if the edge probability between vertex $u$ and $v$ is $P[u,v]$ and independent of other edges. The notation is $A \sim ER(P)$, and the likelihood of $A$ under this model is
	\[\mathcal{L}(A;P) = \prod_{u<v} (P[u,v])^{A[u,v]} (1- P[u,v])^{1-A[u,v]} .\]
\end{definition}
The class label is built into this model as follows: suppose the graph follow ER model conditioned on $Y$, that is
\[ A|Y=y \sim ER(P_y) \qquad \text{ for } y \in\{0,1\}, \]
then vertex $u$ is a signal vertex if and only if $P_{0}[u,v] \neq P_{1}[u,v]$ for some vertex $v$:
\[S = \{u \in V|\exists v \in V,  P_{0}[u,v] \neq P_{1}[u,v]\}.\]
Given this model, the optimal classification performance is achieved by the Bayes classifier $g^*(\cdot)$ \citep{devroye2013probabilistic} defined as
\[ g^*(A) =
\begin{cases}
1       & \quad \text{if } \pi_0 \mathcal{L}  (A;P_{0}) < \pi_1 \mathcal{L}(A;P_{1}),\\
0  & \quad \text{if } \pi_0 \mathcal{L}  (A;P_{0}) \geq \pi_1 \mathcal{L}(A;P_{1}),\\
\end{cases}
\]
where $\pi_0$ and $\pi_1$ are prior probabilities for each class. 

For given sample data $\{(A_i,Y_i), i=1,\ldots,m\}$, these unknown probabilities can be estimated via
\[\hat{\pi}_y =  \frac{\sum_i \mathbb{I}_{\{Y_i=y\}}}{m},\]
\[\hat{P}_y =  \frac{\sum_i \mathbb{I}_{\{Y_i=y\}} A_i}{\sum_i \mathbb{I}_{\{Y_i=y\}}},\]
then the Bayes plug-in classifier $g(\cdot)$ using all vertices is
\[ g(V) =
\begin{cases}
1       & \quad \text{if } \hat{\pi}_0 \mathcal{L}  (A;\hat{P}_0) < \hat{\pi}_1 \mathcal{L}(A;\hat{P}_1),\\
0  & \quad \text{if } \hat{\pi}_0 \mathcal{L}  (A;\hat{P}_0) \geq \hat{\pi}_1 \mathcal{L}(A;\hat{P}_1).\\
\end{cases}
\]
Similarly, the Bayes plug-in classifier $g(\cdot)$ using a set of vertices $U \subset V$ is defined as
\[ g(U) =
\begin{cases}
1       & \quad \text{if } \hat{\pi}_0 \mathcal{L}  (A(U);\hat{P}_0(U)) < \hat{\pi}_1 \mathcal{L}(A(U);\hat{P}_1(U)),\\
0  & \quad \text{if } \hat{\pi}_0 \mathcal{L}  (A(U);\hat{P}_0(U)) \geq \hat{\pi}_1 \mathcal{L}(A(U);\hat{P}_1(U)),\\
\end{cases}
\]
where 
\[\mathcal{L}(A(U);\hat{P}_y(U)) =\prod_{u,v \in U} A[u,v] ^{ \hat{P}_y[u,v]}  (1-A[u,v]) ^{ (1-\hat{P}_y[u,v])}.\]

\section{Theorem Proof}

\subsection{Proof of Theorem~\ref{thm1}}
If two random variables are dependent, their population distance correlation is positive. Therefore, for any subgraph $U$ that includes a signal vertex $u$, there exists a constant $c>0$ such that the population distance correlation between $A(U)[u,\cdot]$ and the label $Y$ is greater than $c$. As $|S|$ is assumed to be fixed, we have
	\[ \min_{u \in S} Dcor(A(U)[u,\cdot],Y) \geq c > 0. \]
Additionally, the class label variable $Y$ and $A(U)[u,\cdot]$ are both bounded because the expected number of edges is bounded. 

We have now met the two requirements to apply Theorem 1 in \citep{li2012feature}. By utilizing the theorem and choosing $\kappa=0$, we can conclude that there exist two positive constants $c_1, c_2$, and for any $0<\gamma < 1/2$ we have
\begin{align*}
\PP(S  \subset \hat{S} )  > 1 - O(p \exp(-c_1 m ^{1-2\gamma}) + mp \exp(-c_2 m^{\gamma})).
\end{align*}
The term $p \exp(-c_1 m ^{1-2\gamma}) + mp \exp(-c_2 m^{\gamma})$ vanishes as $m$ increases to infinity. Therefore, $\PP(S \subset \hat{S} ) \rightarrow 1$ as $m \rightarrow \infty$.

\subsection{Proof of Theorem~\ref{thm2}}
We will establish the ensuing Lemma for the whole graph. The result for $\hat{S}$ immediately follows by substituting the number of edges $e_V$ with $E(\hat{S})$ and adding up the probability error term from the proof of Theorem~\ref{thm1}.
\begin{lemma}
With high probability, $L(g(V)) - L(g^*)$ is bounded by $\epsilon$, that is
\begin{align*}
\PP(L(g(V)) -  L(g^*)  < \epsilon)  \geq 1 - 2(e_{V}+1)\exp \left( \frac{-m c_4 \epsilon^2}{(2e_{V}+\sqrt{2c_4})^2} \right), 
\end{align*}
where $e_{V}$ is the expected number of edges in the whole graph. Moreover, 
\[\EE(L(g(V)))  \leq L(g^*) + \epsilon + 2(e_{V}+1)\exp\left( \frac{-m c_4 \epsilon^2}{(2e_{V}+\sqrt{2c_4})^2}\right).\]
for small $\epsilon > 0$.
\end{lemma}

We first show the Bayes plug-in likelihood $\mathcal{L}(A;\hat{P}_y)$ is close to the true likelihood $\mathcal{L}(A;P_y)$ with high probability. Applying Hoeffding's inequality to $\hat{\pi}_y$, we have
\[ \PP(|\hat{\pi}_y - \pi_y | < \epsilon_1) \geq 1 - 2 \exp(-2m\epsilon_1^2).  \]
By choosing $\epsilon_1$ small enough such that $\hat{\pi}_y > \frac{c_4}{2}$ for some fixed $c_4 > 0$, and applying Hoeffding's inequality to $\hat{P}_{y_{ij}}$, we also have
\[ \PP(|\hat{P}_{y_{ij}} - P_{y_{ij}} | < \epsilon_2) \geq 1- 2 \exp(-m c_4\epsilon_2^2)  .\]
When $|\hat{\pi}_y - \pi_y | < \epsilon_1$ and $|\hat{P}_{y_{ij}} - P_{y_{ij}} | < \epsilon_2$, for any adjacency matrix $A$:
\begin{align*}
	&|\pi_y\mathcal{L}(A;P_y)  -\hat{\pi}_y \mathcal{L}(A;\hat{P}_y)| \\
     &\leq |\pi_y\mathcal{L}(A;P_y) - \pi_y \mathcal{L}(A;\hat{P}_y)| + |\pi_y \mathcal{L}(A;\hat{P}_y) -\hat{\pi}_y \mathcal{L}(A;\hat{P}_y)| \\
	 &< |\pi_y\mathcal{L}(A;P_y) - \pi_y \mathcal{L}(A;\hat{P}_y)|+\epsilon_1 \\
	 &< |\mathcal{L}(A;P_y) - \mathcal{L}(A;\hat{P}_y)|+\epsilon_1 \\
	 &< \epsilon_2 \sum_{i,j}A_{ij}+\epsilon_1.
\end{align*}
The last inequality follows from recursively applying the technique used in the first inequality and the fact that $|\hat{P}_{y_{ij}} - P_{y_{ij}} | < \epsilon_2$. Taking the expectation we have
\begin{align*}
&\EE_{A} (|\pi_0\mathcal{L}(A;P_{0}) -\hat{\pi}_0 \mathcal{L}(A;\hat{P}_0)| + |\pi_1\mathcal{L}(A;P_{1}) -\hat{\pi}_1 \mathcal{L}(A;\hat{P}_1)|) \\  
&\leq \EE_{A}(2\epsilon_2 \sum_{i,j}A_{ij} +2\epsilon_1 ) \\
&\leq 2(e_{V} \epsilon_2+\epsilon_1).
\end{align*}
Setting $ 2(e_{V} \epsilon_2+\epsilon_1)  = \epsilon$ and $2\epsilon_1^2 = c_4 \epsilon_2^2$, we have $\epsilon_2 = \frac{\epsilon}{2e_{V}+\sqrt{2c_4}}$.
Applying Theorem 2.3 in \citep{devroye2013probabilistic} yields
\[\PP(L(g(V)) - L(g^*) < \epsilon) \geq 1 - 2(e_{V}+1)\exp \left( \frac{-m c_4 \epsilon^2}{(2e_{V}+\sqrt{2c_4})^2} \right). \]
We can also further verify that
\begin{align*}
	&\EE(L(g(V)))  - L(g^*) = \EE(L(g(V)) - L(g^*)) \\
	& < \epsilon \mathbb{I}\{L(g(V)) - L(g^*) < \epsilon\}  + \mathbb{I}\{L(g(V)) - L(g^*) \geq \epsilon\} \\
	& < \epsilon + 2(e_{V}+1)\exp \left( \frac{-m c_4 \epsilon^2}{(2e_{V}+\sqrt{2c_4})^2} \right).
\end{align*}

\subsection{Proof of Theorem~\ref{thm3}}
From proof of Theorem~\ref{thm2}, for the whole graph we have 
\begin{align*}
\PP(L(g(V)) - L(g^*) < \epsilon) \geq 1 - 2(e_{V}+1)\exp \left( \frac{-m c_4 \epsilon^2}{(2e_{V}+\sqrt{2c_4})^2} \right).
\end{align*}
To achieve asymptotically optimal classification of the whole graph, i.e., $L(g(V)) \rightarrow L(g^*)$, it suffices for the second term to approach $0$, which happens when $e_{V}=o(m^{\frac{1}{2}})$. Conversely, if we have a graph where $e_{V} = c n^2$ for some positive constant $c \in (0,1]$ and $n=O(m^{\frac{1}{4}})$, the second term no longer approaches zero, leading to worse-than-optimal classification. 

For the estimated signal subgraph we have
\begin{align*}
\PP(L(g(\hat{S})) - L(g^*) < \epsilon) &\geq 1 - 2(E(\hat{S})+1)\exp \left( \frac{-m c_4 \epsilon^2}{(2E(\hat{S})+\sqrt{2c_4})^2} \right)\\
	& \qquad - c_3 (p \exp(-c_1 m ^{\frac{1}{3}}) + mp \exp(-c_2 m ^{\frac{1}{3}})).
\end{align*}
For $L(g(\hat{S})) \rightarrow L(g^*)$, it suffices for the second and third terms to converge to $0$. As $|\hat{S}|$ is assumed bounded, so is $E(\hat{S})$. Thus the second term vanishes as $m \rightarrow \infty$, so is the third term. 

Therefore, $L(g(\hat{S})) \rightarrow L(g^*)$. When $n=O(m^{\frac{1}{4}})$, $L(g(V)) \not\rightarrow L(g^*)$, such that $L(g(\hat{S})) < L(g(V))$ for sufficiently large $n,m$. 

\end{document}